# A non-algorithmic approach to “programming” quantum computers via machine learning

Nathan Thompson
*Dept. of Math. Stat. and Phys.*
*Wichita State University*
Wichita, KS USA
thompson@math.wichita.edu

James Steck
*Dept. of Aerospace Engineering*
*Wichita State University*
Wichita, KS, USA
james.steck@wichita.edu

Elizabeth Behrman
*Dept. of Math. Stat. and Phys.*
*Wichita State University*
Wichita, KS, USA
behrman@math.wichita.edu

***Abstract*— Major obstacles remain to the implementation of macroscopic quantum computing: hardware problems of noise, decoherence, and scaling; software problems of error correction; and, most important, algorithm construction. Finding truly quantum algorithms is quite difficult, and many of these genuine quantum algorithms, like Shor’s prime factoring or phase estimation, require extremely long circuit depth for any practical application, which necessitates error correction. In contrast, we show that machine learning can be used as a systematic method to construct algorithms, that is, to non-algorithmically “program” quantum computers. Quantum machine learning enables us to perform computations without breaking down an algorithm into its gate “building blocks”, eliminating that difficult step and potentially increasing efficiency by simplifying and reducing unnecessary complexity. In addition, our non-algorithmic machine learning approach is robust to both noise and to decoherence, which is ideal for running on inherently noisy NISQ devices which are limited in the number of qubits available for error correction. We demonstrate this using a fundamentally non-classical calculation: experimentally estimating the entanglement of an unknown quantum state. Results from this have been successfully ported to the IBM hardware and trained using a hybrid reinforcement learning method.**

## I. Introduction

For several decades now the prospect of macroscopic quantum computers, able to solve large classes of difficult problems, has been “ten years away.” We do have thousand-qubit size “quantum annealing” machines [1], to solve optimization problems through adiabatic evolution to the ground state of a designed Hamiltonian, but programmable quantum computers remain small and their applicability limited. One major obstacle is the construction of algorithms that take advantage of the fundamental quantum nature of reality. There are still only a very few. Most fall into one of three categories: those using a quantum Fourier transform, like Shor's algorithm [2]; those using amplitude amplification, like Grover's algorithm [3]; and those using quantum walks [4]. Speedup varies; Shor's, and some quantum walk algorithms, provide an exponential advantage over the best known classical algorithm in each case, but the speedup with Grover is only quadratic. We do not yet know whether there exists any quantum advantage for broad classes of problems [5] [6], much less, what it will be in each case. Nor do we have a general process to factor an arbitrary N-qubit unitary efficiently to generate the quantum machine language necessary, in the case of the gate model; or to design a Hamiltonian whose ground state will be the answer to an optimization problem, in the case of quantum annealing.

Since the 1990s, our research group has been investigating the advantages of a marriage of machine learning and quantum computing to answer this need [7] [8] [9]. The basic idea is that a quantum system can itself act as a neural network: The state of the system at the initial time is the “input”; a measurement (observable) on the system at the final time is the “output”; the states of the system at intermediate times are the hidden layers of the network. If we know enough about the computation desired to be able to construct a comprehensive set of input-output pairs from which the net can generalize, then, we can use techniques of machine learning to bypass the algorithm-construction problem. Moreover, this approach is scalable [10] as we are able to train iteratively in larger and larger stages and use knowledge of a smaller system to make systematic inferences about a larger one. In addition, our method promises to be generally robust to both noise and to decoherence [11] [12]. Machine learning may also be helpful in the factorization problem [13], and in the Hamiltonian design problem [14].

Entanglement estimation is a good example of a nontrivial, intrinsically quantum mechanical, calculation for which we have no general algorithm [15] [16]. Indeed, it has been shown that the quantum separability problem (determination of entanglement) is NP-hard [17]. In previous work we succeeded in mapping a function of a measurement at the final time to a witness of the entanglement of a two-qubit system in its initial state [9]. The “output” (result of the measurement of the witness at the final time) will change depending on the time evolution of the system, which is of course controlled by the Hamiltonian: by the tunneling amplitudes $\{K\}$, the qubit biases $\{\varepsilon\}$, and the qubit-qubit coupling $\zeta$. Thus we can consider these functions $\{K_A, K_B, \varepsilon_A, \varepsilon_B, \zeta\}$ to be the “weights” to be trained. (Here A and B refer to the two qubits.) We then use a quantum version [9] of backpropagation [18] to find optimal functions such that our desired mapping is achieved. Full details are provided in [8] [9]. From a training set of only four pure states, our quantum neural network successfully generalized the witness to large classes of states, mixed as well as pure [10]. Qualitatively, what we are doing is using machine learning techniques to find a "best" hyperplane to divide separable states from entangled ones, in the Hilbert space.

Of course, this method is necessarily "off-line" training, since it is not possible to do backpropagation without knowing the state of the system at intermediate times (in the hidden layers); quantum mechanically, this is impossible without collapsing the wavefunction and thereby destroying the superposition, which rather obviates the whole purpose of doing quantum computation. That is, quantum backpropagation can only be done on an (auxiliary) classical computer, simulating the quantum computer, and this simulation will necessarily contain uncertainties and errors in modeling the behavior of the actual quantum computer. The results from off-line quantum backprop, can, of course, be used as a good starting point for true online quantum learning, where reinforcement learning is used to correct for uncertainty, noise, and decoherence in the actual hardware of the quantum computer. Here, we present such a method, port it to the IBM hardware [19], and demonstrate its effectiveness.

The next section is a brief overview of the theory of our Quantum Backprop technique, followed by Section III which contains the details of our Reinforcement Learning alternative for on-line learning with quantum hardware. In Section IV we port the method to Qiskit. We conclude in Section V.

## II. Dynamic Learning Quantum Backprop

The density matrix, ρ, of a quantum system as a function of time obeys the Schrödinger equation $\frac{d\rho}{dt} = \frac{1}{i\hbar}[H, \rho]$ , where H is the Hamiltonian and ħ is Planck's constant divided by 2π. The formal solution of the equation is $\rho(t) = e^{i\mathcal{L}t}\rho(t_0)$ where $\mathcal{L}$ is the Liouville operator. Thus, the time evolution equation for the density matrix maps the initial state $\rho(t_0)$ (input data for the quantum computer) to the final state $\rho(t_f)$ (output). The mapping is accomplished by the exponential of the Liouville operator, $e^{i\mathcal{L}t}$. Parameters in the system Hamiltonian H are physical interactions and fields in quantum hardware and can be adjusted experimentally as functions of time. "Programming" this quantum computer involves finding the parameters using machine learning that yield the desired computation. Thus we can train the system to evolve in time initial (input) to target final (output) states; yielding a quantum system that accurately approximates a chosen function, such as logic gates, benchmark classification problems, or, since the time evolution is quantum mechanical, a quantum function like entanglement.

If we think of the time evolution operator in terms of the Feynman path integral picture [20], the system can be seen as analogous to a neural network, yet quantum mechanical. That is, instantaneous values taken by the quantum system at intermediate times, which are integrated over, play the role of "virtual neurons" [9]. In fact, this system is a deep learning system, as the time dimension controls the propagation of information from the input to the output of the quantum system, and the depth is controlled by how finely the parameters are allowed to vary with time. The real time evolution of a multi-qubit system can be treated as a neural network, because its evolution is a nonlinear function of the various adjustable parameters (weights) of the Hamiltonian. The goal of learning as applied to this quantum system is to "program" the system via adjusting the external parameters to force it to calculate target outputs in response to given inputs. This is done via a neural network supervised learning paradigm which we have extended to the quantum system. The method, derived below, follows the methodology of Yann LeCun's Lagrangian formulation derivation of backpropagation [21] and Paul Werbos's description of backpropagation through time [18], and follows some of our earlier work [22] [23] on learning in non-linear optical materials and in training of quantum Hopfield networks. For as long as coherence can be maintained experimentally, it is a quantum neural network (QNN).

Our learning rule for the quantum system based on dynamic backpropagation is derived as follows. Given an input (initial density matrix), $\rho_0$, and a target output, d (a "training pair"), we develop a weight update rule based on gradient descent to adjust the system parameters, i.e., train the system "weights", to reduce the squared error between the target, d, and the output, Output. While minimizing the squared error, the system's density matrix, ρ(t), is constrained to satisfy the Schrödinger equation for all time in the interval $(t_0, t_f)$.

We define a Lagrangian, L, to be minimized, as

$$L = \frac{1}{2}[d - < O(t_f) >]^2 + \int_{t_0}^{t_f} \lambda^+(t)\left(\frac{\partial\rho}{\partial t} + \frac{i}{\hbar}[H,\rho]\right)\gamma(t) \quad dt$$

where the Lagrange multiplier vectors are $\lambda^+(t)$ and γ(t) (row and column, respectively), and $O$ is an output measure (or some function of a measure), which is chosen for the particular problem under consideration. As an example, for our entanglement witness application, we defined the output as:

$$\langle O(t_f)\rangle = tr[\rho(t_f)O] = \sum_i p_i \langle\psi_i(t_f)|O|\psi_i(t_f)\rangle$$

where *tr* stands for the trace of the matrix, and where the density matrix is represented in terms of the chosen basis as $\rho = \sum_i p_i |\psi_i\rangle\langle\psi_i|$. We take the first variation of $L$ with respect to ρ, set it equal to zero, then integrate by parts to give the following equation which can be used to calculate the vector elements of the Lagrange multipliers ("error deltas" in neural network terminology) that will be used in the learning rule: $\gamma_i \frac{\partial\gamma_j}{\partial t} + \frac{\partial\lambda_i}{\partial t}\gamma_j - \frac{i}{\hbar}\sum_k \lambda_k H_{ki}\gamma_j + \frac{i}{\hbar}\sum_k \lambda_i H_{jk}\gamma_k = 0$ , which is solved backward in time, with the boundary conditions at the final time $t_f$ given by $-[d - < O(t_f) >]O_{ji} + \lambda_i(t_f)\gamma_j(t_f) = 0$. The gradient descent rule to minimize L with respect to $w$ is $w_{new} = w_{old} - \eta\frac{\partial L}{\partial w}$ for each "weight" parameter $w$, where η is the learning rate, and where the derivative is given by

$$\frac{\partial L}{\partial w} = \frac{i}{\hbar}\int_0^{t_f} \lambda^+(t)\left[\frac{\partial H}{\partial w},\rho\right]\gamma(t)dt$$

$$= \frac{i}{\hbar}\int_0^{t_f} \sum_{ijk}\left(\lambda_i(t)\frac{\partial H}{\partial w}\rho_{kj}\gamma_j - \lambda_i(t)\rho_{ik}\frac{\partial H_{kj}}{\partial w}\gamma_j\right)dt$$

The above technique, since it uses the density matrix, is applicable to any state of the quantum system, pure or mixed.

## III. REINFORCEMENT LEARNING

### A. Reinforcement Learning of the Fourier Quantum Parameters

Each of the quantum system parameters can vary with time as described earlier. For reinforcement learning, each quantum parameter is represented as a Fourier expansion in time:

$$P(t) = P_0 + \sum_{i=1}^{n}\left[S_n \sin\left(n\frac{\pi}{T}t\right) + C_n \cos\left(n\frac{\pi}{T}t\right)\right] \quad (1.1)$$

where T is the end quantum system simulation time where the output measures are taken. This gives a limited population of Fourier coefficients to vary during the reinforcement learning and is motivated by the results shown in [11] where backpropagation is used to train the quantum parameters allowing any continuous functions of time but the resulting parameters have obvious simple frequency content. In this paper we showed that fitting the parameters with Fourier series for sine and cosine gave equivalent computing results.

Learning of each of the parameters is done via a hybrid method which uses small variations of the Fourier coefficients to calculate the gradient of the output error which is then used is a straightforward gradient descent learning rule.

For a given training pair in the training set, the quantum system is presented with the input, the system runs (with the current parameters calculated from the current Fourier coefficients) until the final time $t_f$ where the output is calculated via the output measures on the final state. The output is compared to the target value and an output error is calculated, $E_{nom}$. In the backprop method, this output error is then backpropagated via quantum backprop to calculate gradients at each time step. In the hybrid reinforcement learning the following happens.

Choosing a single quantum parameter and a single Fourier parameter in 1.1, this parameter is varied by a small amount. For example, for P0 the new parameter would be given by

$$P_{0,new} = P_0 + \Delta P_0 \quad (1.2)$$

The quantum system is again presented with the input; the system then runs with the parameters calculated using the modified Fourier coefficients; the output is calculated; an output error Emod is calculated and compared to the error Enom; a gradient is calculated

$$Gradient = \frac{E_{mod} - E_{nom}}{\Delta P_0}; \quad (1.3)$$

and, finally, this gradient is used to update the parameter using a specified learning rate $\eta_{P_0}$ via

$$P_0 = P_0 - \eta_{P_0} Gradient \quad (1.4)$$

This is repeated for each Fourier coefficient and each quantum parameter, using the same input and target output. Each successive training pair is then processed in the same way until the entire list of training pairs is exhausted, constituting one epoch of training.

### B. Reinforcement Learning Results

Matlab code implements the learning algorithm above and calls a Simulink simulation of the quantum system. Compared to the quantum backprop method, reinforcement learning, in simulation, takes about from 25 times more computation time. The tunneling frequency is initialized to $2.5\text{x}10^{-3}$ GHZ, is changed by .02% to calculate the gradient and a learning rate of 0.00000002 is used. The bias is initialized to $1.0\text{x}10^{-4}$ GHZ, is changed by .02% to calculate the gradient and a learning rate of zero (not trained) is used. The qubit coupling matrix off-diagonal elements representing qubit to qubit coupling are initialized to $1.0\text{x}10^{-4}$ GHZ, is changed by .02% to calculate the gradient and a learning rate of 0.0000004 is used. The on-diagonal coupling of a qubit to itself is, of course, zero. The entanglement witness calculation described above is the quantum "program" to be learned. Three Fourier parameters in equation (1.1) are used, that is n=3. Systems with 2, 3, 4 and 5 qubits are trained. A plot of the RMS error vs epoch is shown as well as plots of how each quantum parameter varies with time after training is completed.

Results for the 2-qubit system are shown in Figures 1 through 3, using reinforcement learning, and in Figures 4 through 6, using our earlier quantum backprop technique.

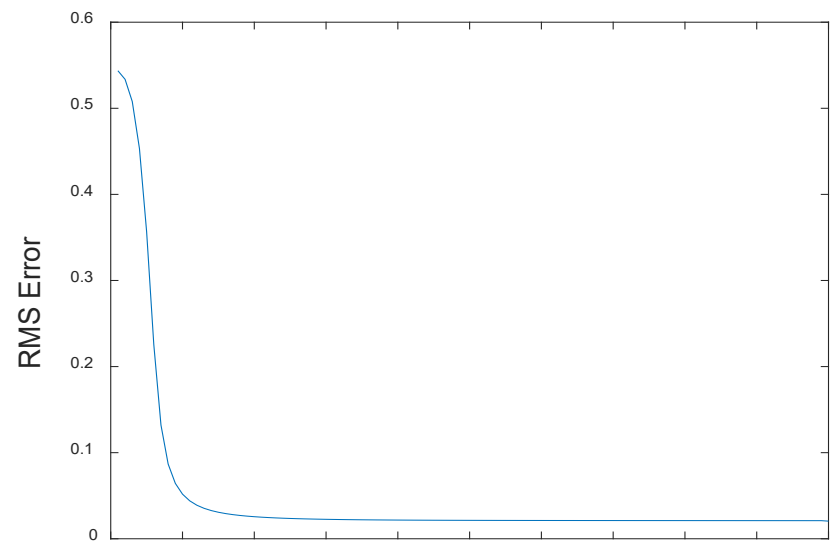

Fig. 1. RMS error vs Epoch for 2 qubit entanglement witness training

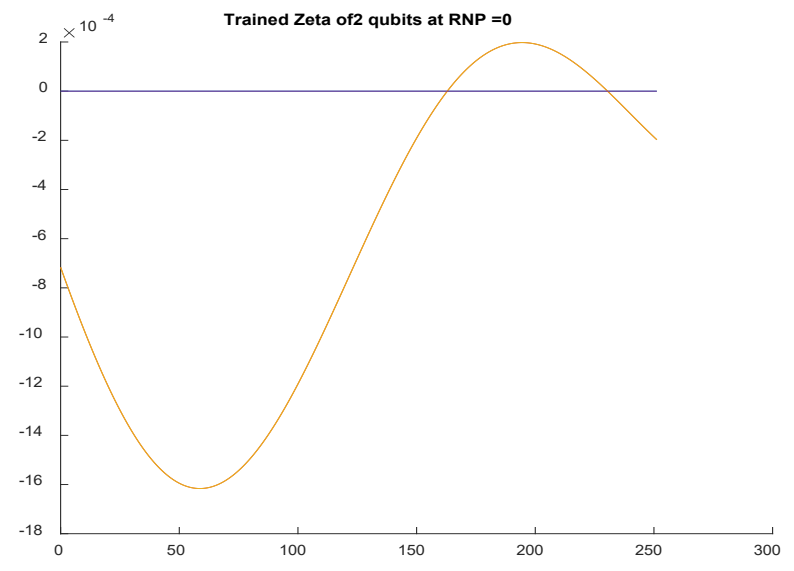

Fig. 2. Coupling parameter ζ as a function of time for 2 qubit entanglement witness training

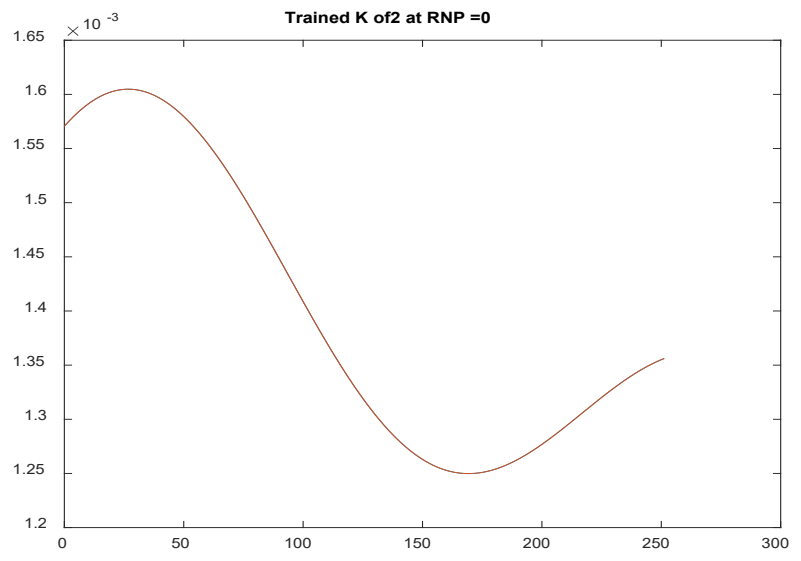

Fig. 3. Tunneling parameter K as a function of time for 2 qubit entanglement witness

Comparing Figures 1 through 3 to Figures 4 through 6, we see that the reinforcement methods give a different set of quantum parameter functions, but the RMS errors are similar. Testing of each also produces equivalent results. Clearly, the set of functions that produces this entanglement witness is not unique; this is unsurprising as we have seen similar behavior in earlier work with entanglement witnesses [11]. While backprop is computationally much faster in simulation, it cannot be implemented on the IBM hardware, while the reinforcement learning method can.

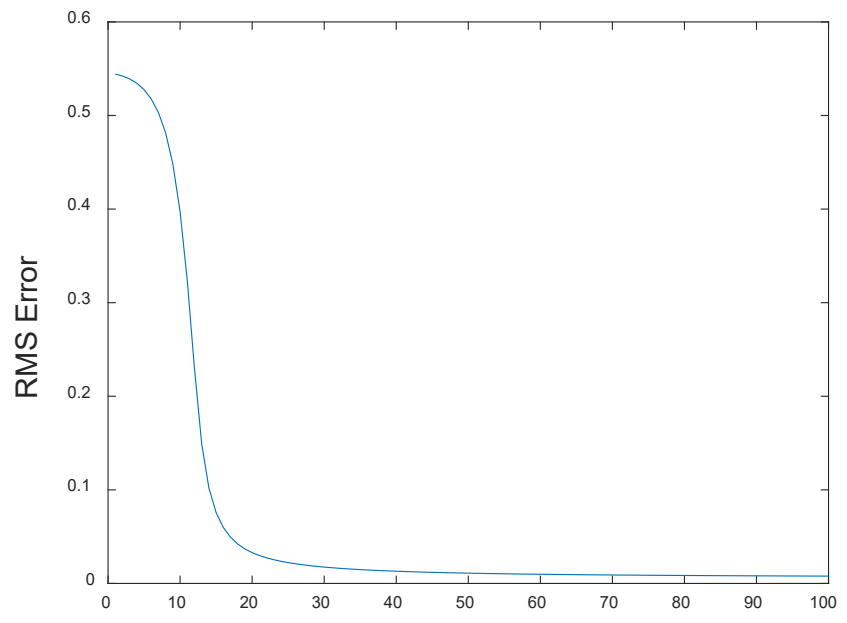

Fig. 4. RMS error vs Epoch for 2 qubit entanglement witness **backprop training**

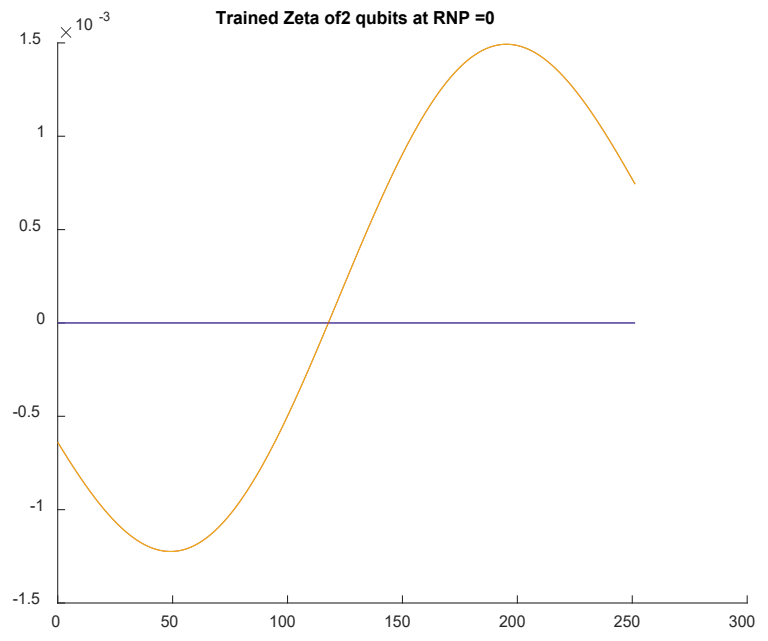

Fig. 5. Coupling parameter ζ vs time for 2 qubit enntanglement witness **backprop training**

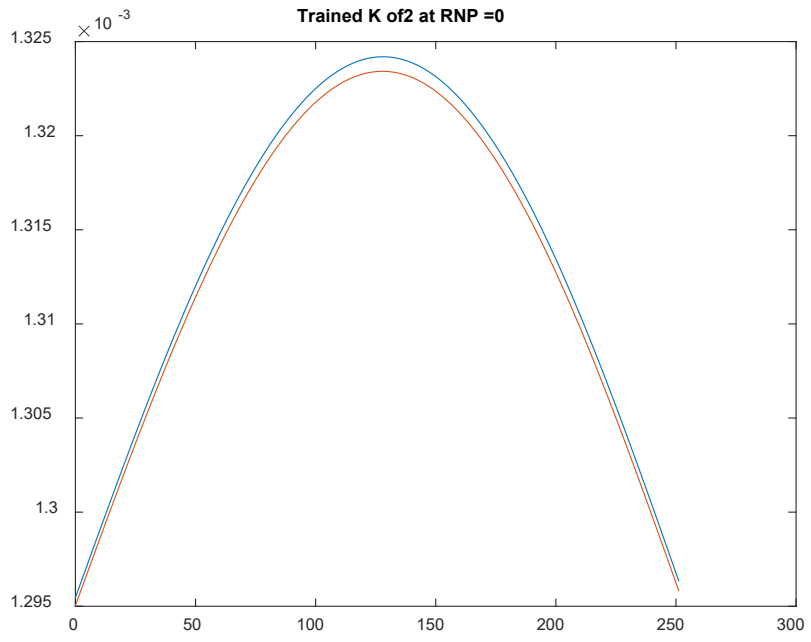

Fig. 6. Tunneling amplitude K for 2 qubit entanglement witness **backprop training**

We now extend our results from the two-qubit system to a many-qubit system, using a method [10] [24] we call “iterative staging”, because we use knowledge of a smaller system to infer partial knowledge of a larger. Here, we use the trained parameters for the 2-qubit system to initialize a 3-qubit system as a starting point. The tunneling parameters are simply copied to all three qubits. The coupling ζ between the 2 qubits is copied onto the 3-qubit system to be the initial value for each of the three pairwise couplings. In prior work, we discovered that this initialization significantly reduces the number of epochs required to train systems with higher numbers of qubits [10].

Reinforcement learning for 3 qubits results are shown in Figures 7 through 9. Again, we see that the error rapidly decreases and the resulting parameters are a good “program” for the calculation.

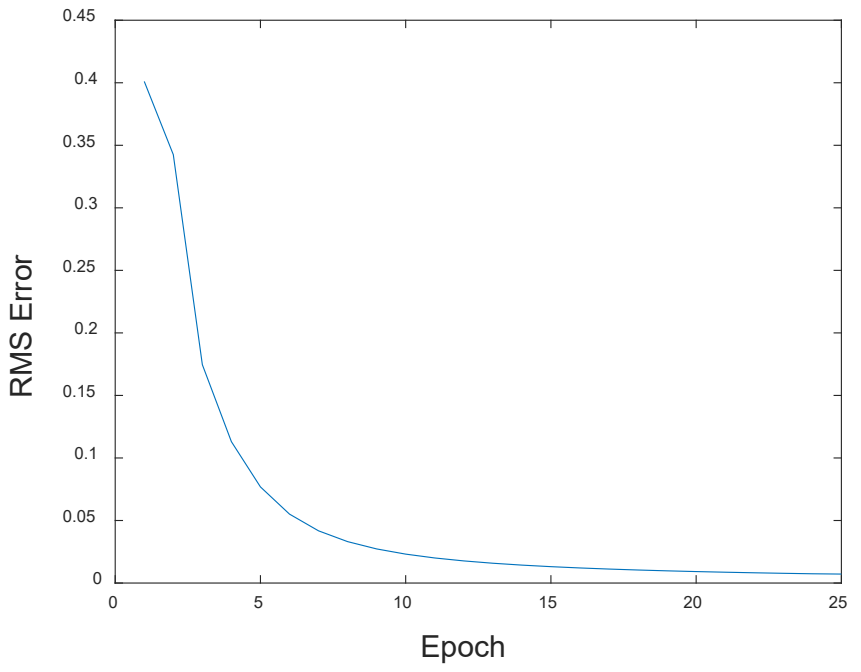

Fig. 7. RMS error vs Epoch for 3 qubit entanglement witness training

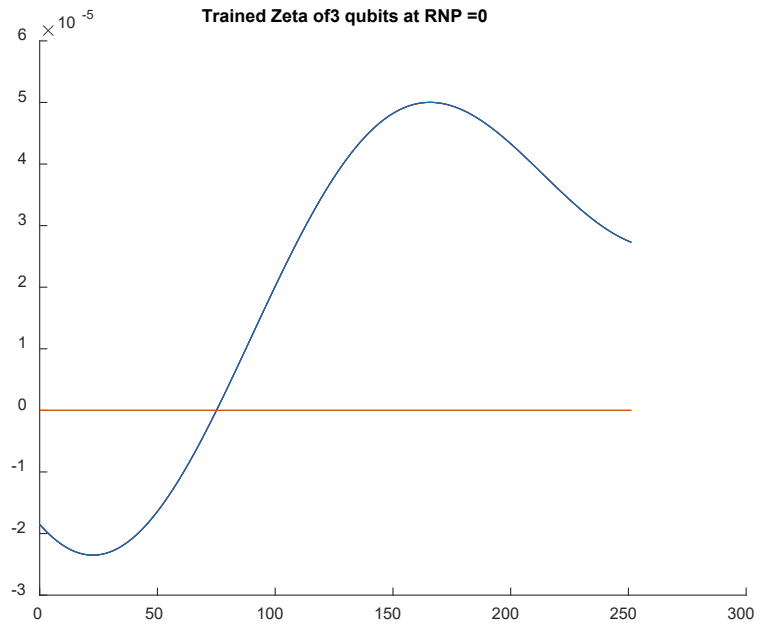

Fig. 8. Coupling parameter ζ as a function of time for 3 qubit entanglement witness

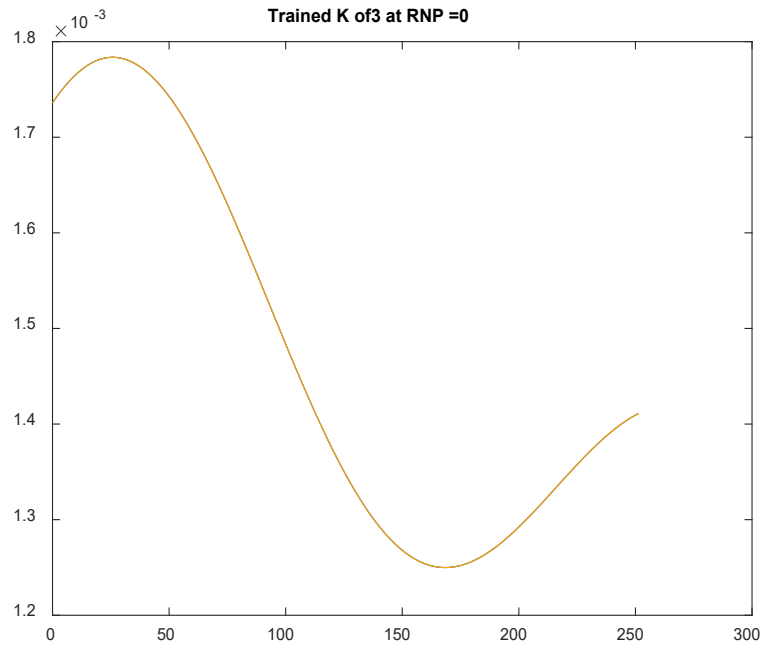

Fig. 9. Tunneling parameter K as a function of time for 3 qubit entanglement witness

For the 4-qubit system initial parameters, we copy the trained results from the 3-qubit system. Results are shown in Figures 10 through 12.

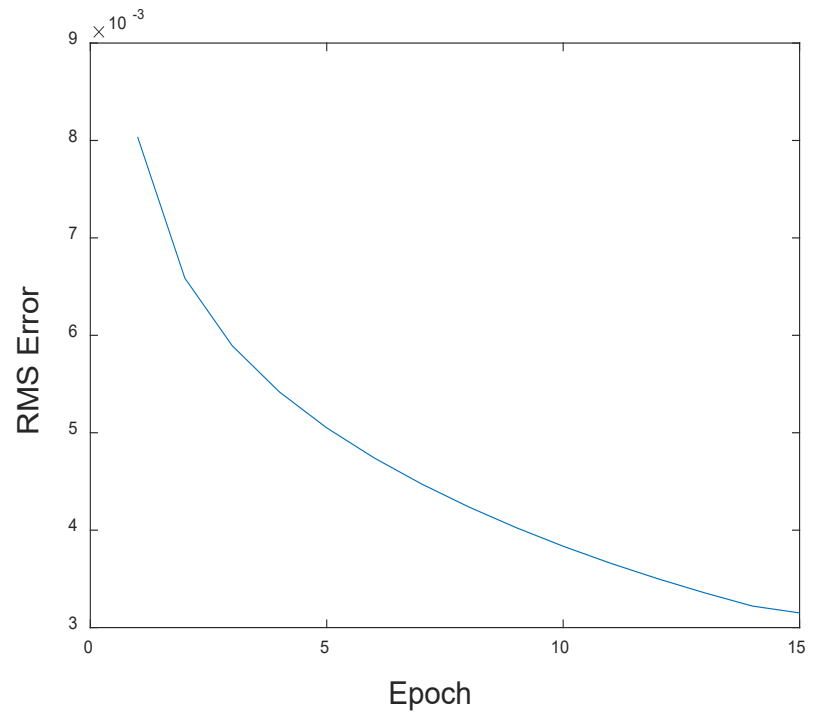

Fig. 10. RMS error vs Epoch for 4 qubit entanglement witness training

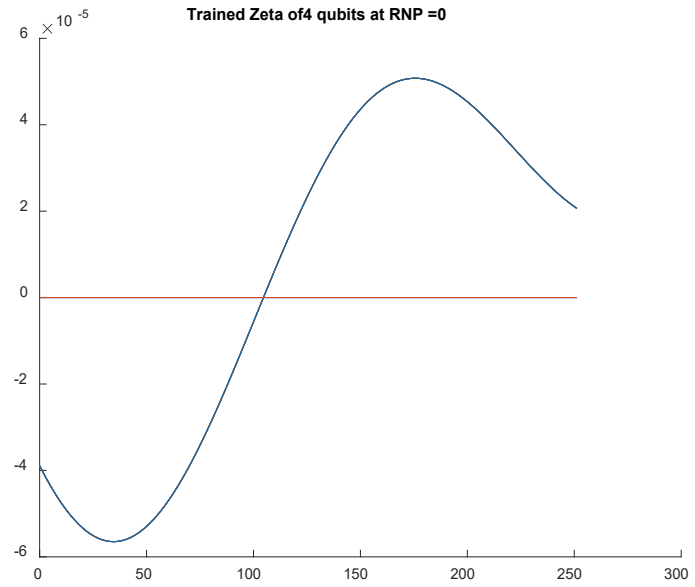

Fig. 11. Coupling parameter ζ as a function of time for 4 qubit entanglement witness

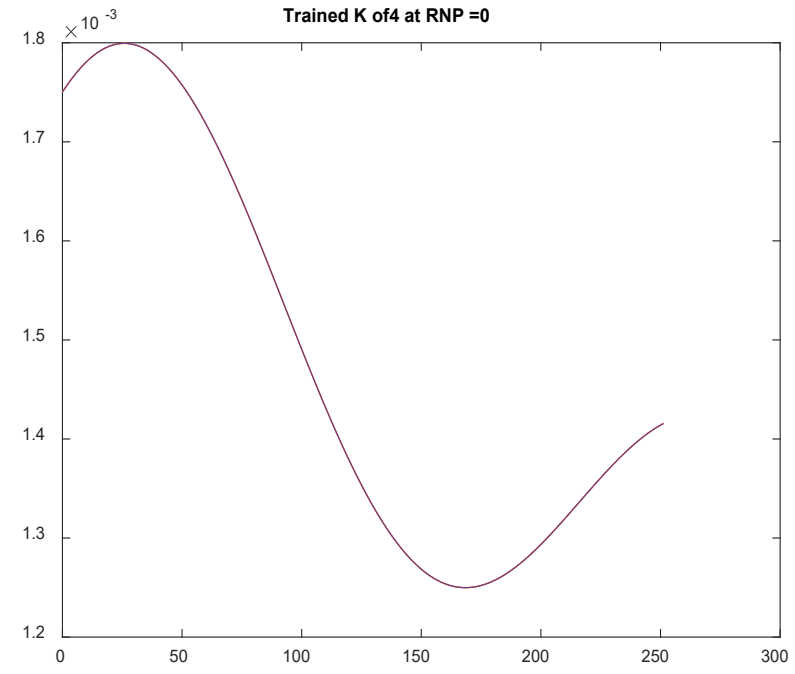

Fig. 12. Tunneling parameter K as a function of time for 4 qubit entanglement witness

For the 5-qubit system initial parameters we copy the trained results from the 4-qubit system. Results are shown in Figures 13 through 15.

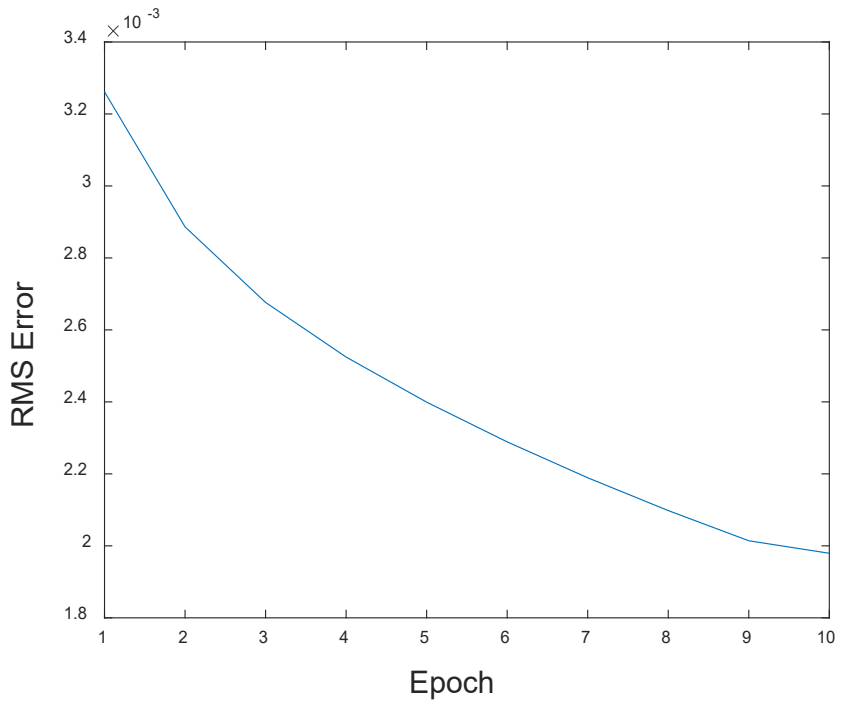

Fig. 13. RMS error vs Epoch for 5 qubit entanglement witness training

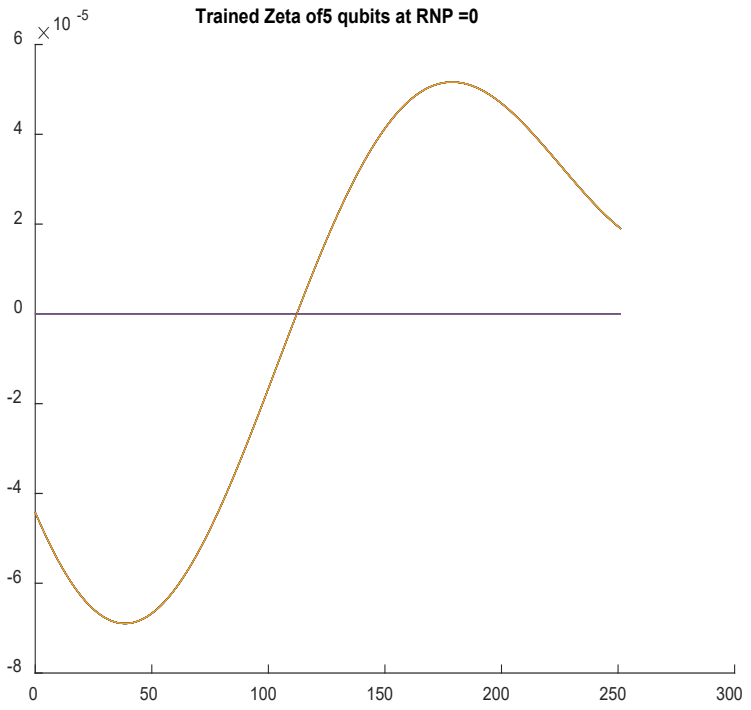

Fig. 14. Coupling parameter ζ as a function of time for 5 qubit entanglement witness

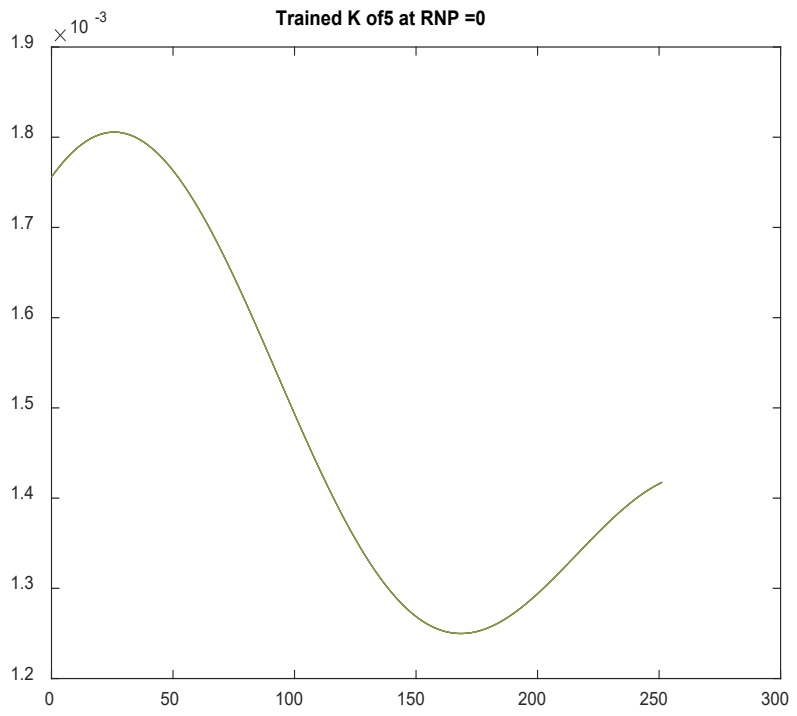

Fig. 15. Tunneling parameter K as a function of time for 5 qubit entanglement witness

## IV. Implementation of Reinforcement Learning in IBM Qiskit

Implementing the pairwise entanglement witness in the IBM hardware requires some modifications. The IBM Qiskit library utilizes a quantum gate model, so we must use a gate representation of the operator. The witness is constructed by first approximating the values of the tunneling, bias, and coupling parameters as piecewise constant, where the total evolution time is divided into 4 segments. These piecewise constant parameters are used to form the Hamiltonian for the time evolution operator, which is converted into a sequence of gates, a quantum circuit. Full details of the gate representation conversion and a comparison with the behavior of the continuum parameters are presented in [24].

The circuit representation results in 20 independent weights $w_j$ for the entanglement witness. For reinforcement learning, the training process is very similar to the Matlab implementation, with necessary changes for the IBM system. First, one of the training states is prepared, then the entanglement witness is applied to it and an expectation value for the witness is returned. Using the current weights, expectation values for each state in the training set are computed and subtracted from the target values to generate a RMS difference output error $E_{nom}$. A single weight $w_j$ is adjusted by a small amount as in (1.2), and the output error is then computed with the modified $w_j$, yielding $E_{mod}$. Equations (1.3) and (1.4) are used to update $w_j$ according to the specified learning rate $\eta_{wj}$, and the process is repeated for each of the 20 weights, constituting one epoch of training.

For training, the tunneling amplitude K was initialized to $2.0\text{x}10^{-3}$ GHZ, bias ε initialized to $1.0\text{x}10^{-4}$ GHZ, and coupling ζ initialized to $1.0\text{x}10^{-4}$ GHZ, for all time segments. The learning rate of the tunneling was $1.0\text{x}10^{-2}$ and the other learning rates were set to $1.0\text{x}10^{-3}$ since experimentation revealed that the system was the most sensitive to changes in the tunneling parameter. Training was successful, but improvement stopped after approximately 2000 epochs where the RMS oscillated near 0.02.

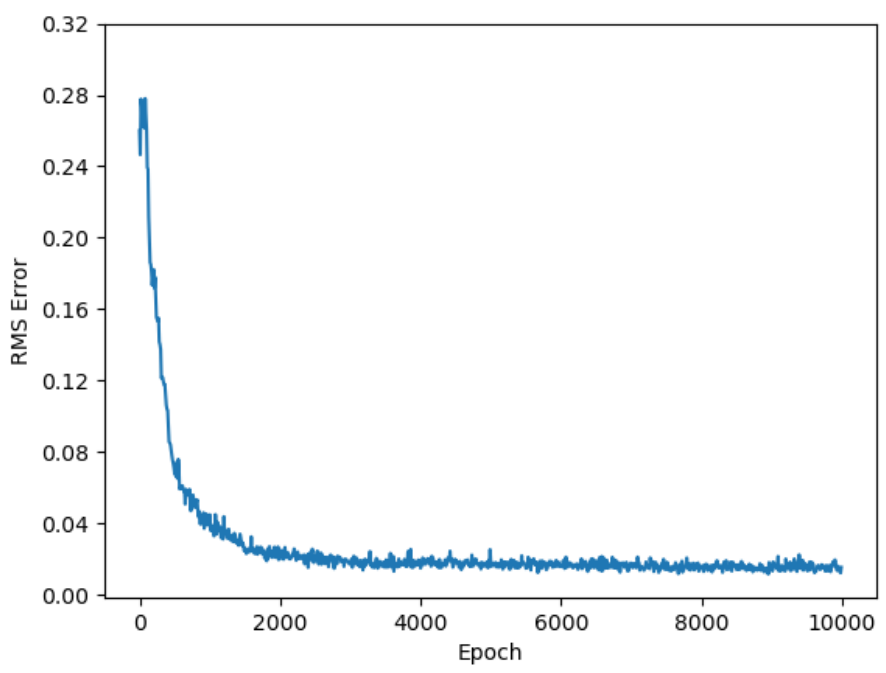

Fig. 16. RMS Error for training 2 qubits in IBM Qiskit system

Parameter training for the tunneling, bias, and coupling all appear to have been completed due to each parameter set aside from the coupling showing clear asymptotic behavior.

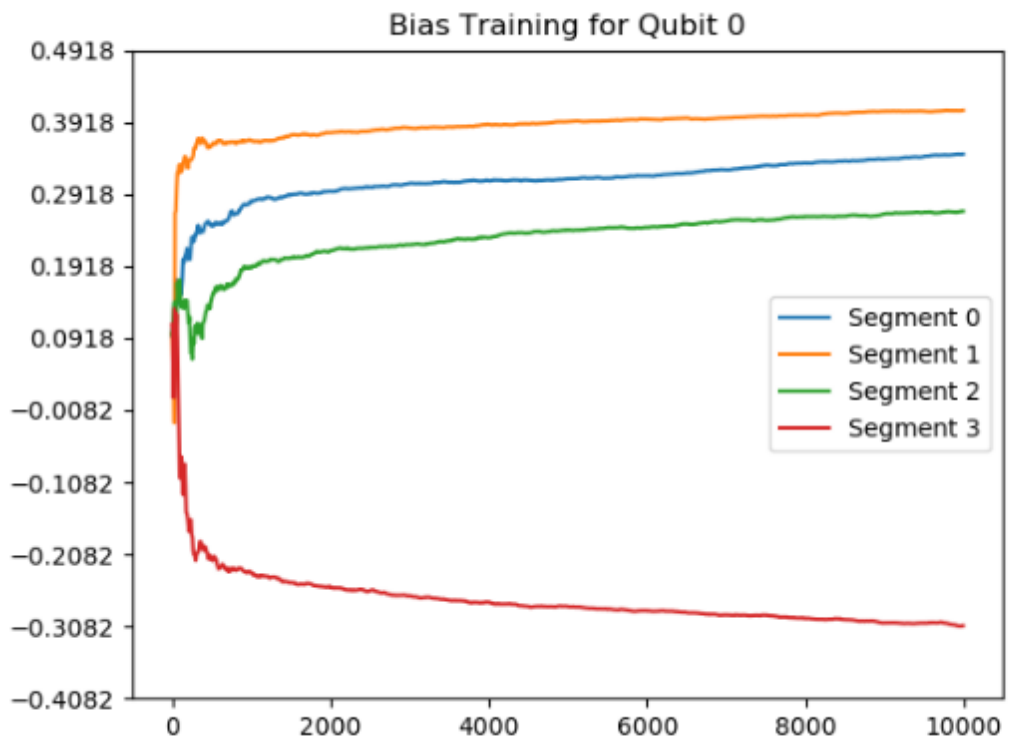

Fig. 17. Bias parameter for the first qubit as a function of time training 2 qubits in IBM Qiskit system

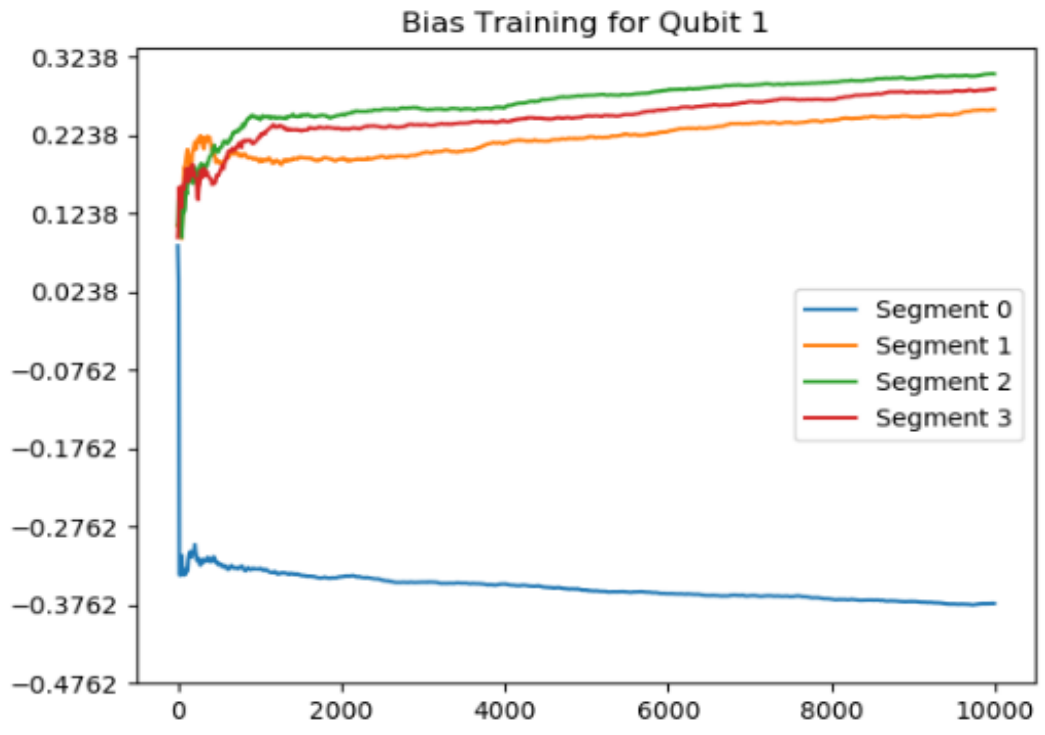

Fig. 18. Bias parameter for the second qubit as a function of time training 2 qubits in IBM Qiskit system

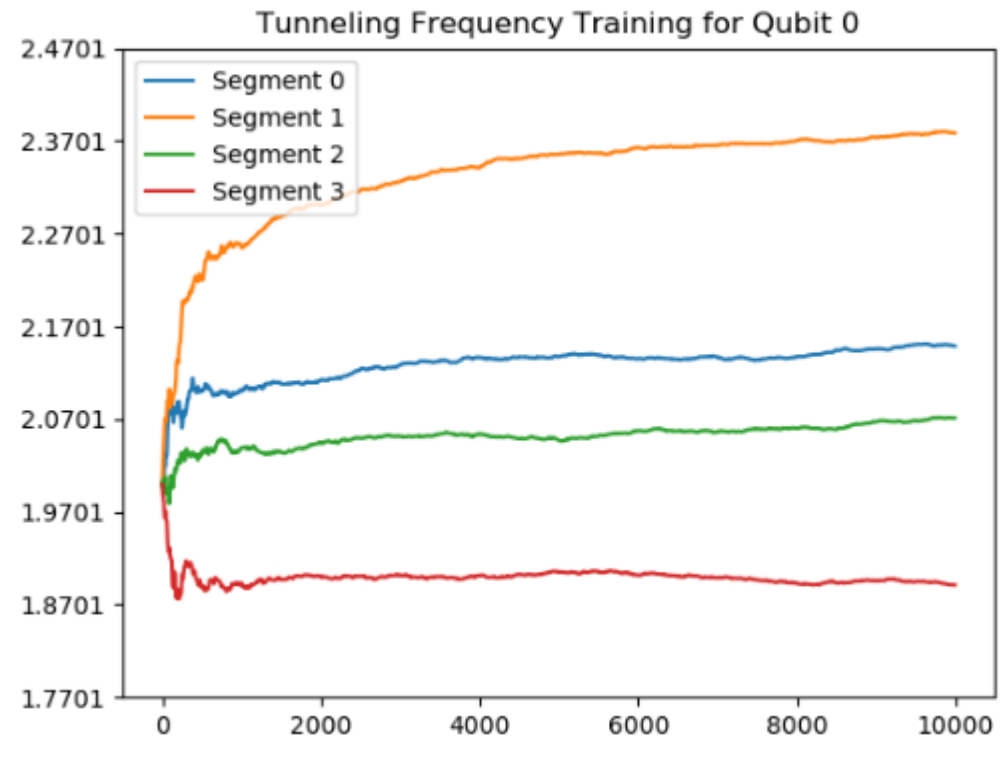

Fig. 19. Tunneling parameter for the first qubit as a function of time training 2 qubits in IBM Qiskit system

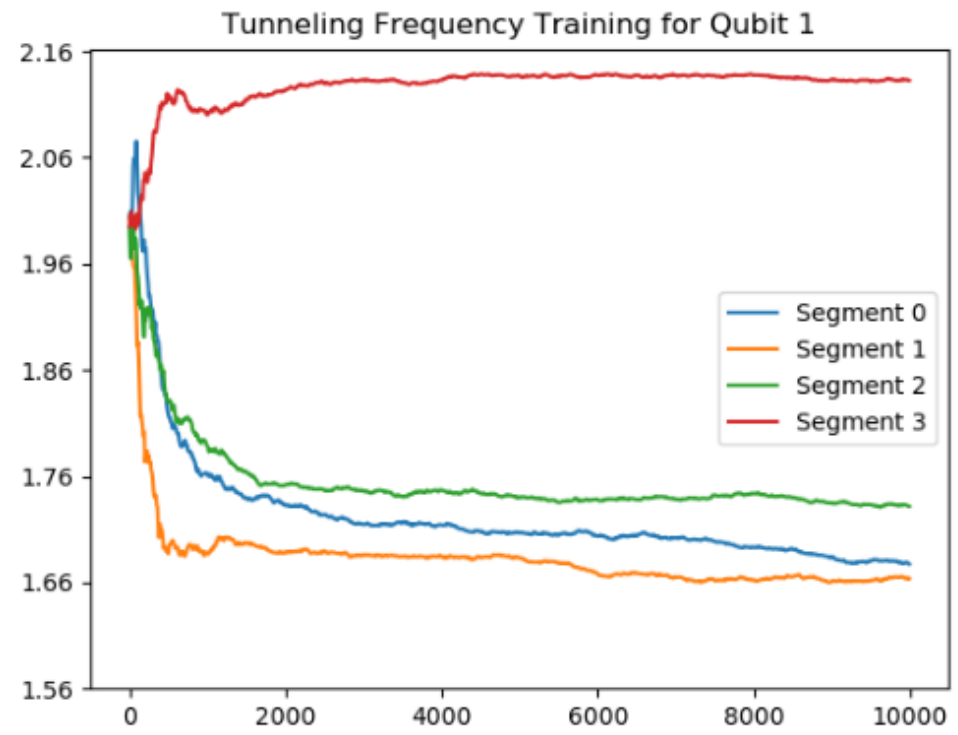

Fig. 20. Tunneling parameter for the second qubit as a function of time training 2 qubits in IBM Qiskit system

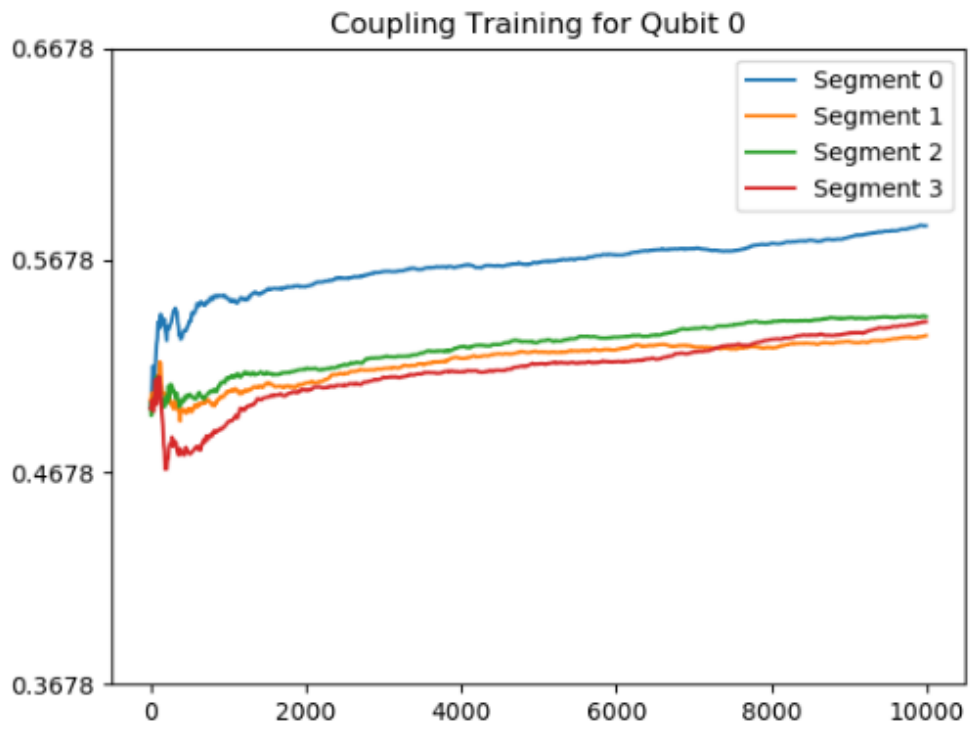

Fig. 21. Inter qubit coupling parameter as a function of time training 2 qubits in IBM Qiskit system

Again, we see that in this high dimensional parameter space the set of parameter functions that succeeds in the mapping we desire, is not unique. Of course, it would be very surprising if it were. The training methods are different. In Qiskit by necessity we used piecewise constant parameter functions instead of Fourier series; learning rates needed were very different; the functions produced were asymmetric; much noisier training meant training was much more computationally intensive.

The quantum backprop and Matlab reinforcement learning produce symmetrical parameter functions for the two qubits: that is, the KA and KB functions are the same. This is probably related to the fact that we did not need to train the bias functions for our results; in the Qiskit runs, once the symmetry is (perhaps randomly) broken, there is no natural way to regain the symmetry, and the function evolution behaves rather as if it were on an unstable equilibrium. We also notice that training is much more difficult in the quantum simulator: it takes an order of magnitude more epochs to reduce the error to the single-percent range (where it is within machine error.) Also, the optimal learning rates (found by trial and error) were very different for the different methods.

## V. Discussion and conclusions

The major contribution of this paper is the demonstration of the feasibility of true online training of a quantum system to do a quantum calculation. It is a well-known theorem that a very small set of gates (e.g., the set {H, T, S, CNOT}) is universal. This means that any n-qubit unitary operation can be approximated to an arbitrary precision by a sequence of gates from that set. But there are many calculations we might like to do, for which we do not know an optimal sequence to use, or even, perhaps, any sequence to use. And there are many questions we might want to answer for which we do not even have a unitary, that is, an algorithm. Calculation of entanglement of an N-qubit system is a good example of such a question: we do not have a general closed form solution, much less know an optimal set of measurements to make on a system whose density matrix is unknown, to determine its entanglement.

Quantum machine learning methods like the ones used here are systematic methods for dealing with these problems. Here we show that they are in fact directly implementable on existing hardware. Our iterative staging technique makes scaleup relatively easy, as most of the training for a system of N qubits has already been accomplished in the system for (N-1) qubits. And while training on actual quantum hardware does prove somewhat more challenging, that is all the more reason for a machine learning approach: in any physical implementation there are always sources of error that in general are unknown (interactions, flaws, incomplete and damaged data). With machine learning we can deal with all these problems automatically.

## VI. Acknowledgment

We all thank the entire research group for helpful discussions: Nam Nguyen, Saideep Nannapaneni, William Ingle, Henry Elliott, Ricardo Rodriguez, and Sima Borujeni.